\documentclass[%
 reprint,
superscriptaddress,
nofootinbib,
 amsmath,amssymb,
 aps,
]{revtex4-2}
\usepackage{graphicx}
\usepackage{dcolumn}
\usepackage{bm}
\usepackage[version=4]{mhchem} 
\usepackage{subfig}
\usepackage{float}
\usepackage{caption}
\usepackage{multirow}
\usepackage{array}
\usepackage{booktabs}
\usepackage[colorlinks,
            linkcolor=red,
            anchorcolor=blue,
            citecolor=blue
           ]{hyperref}

\graphicspath{ {figure/} }
\begin{document}

\title{Deciphering the spectral bumps of Galactic cosmic rays through gamma-ray observations of nearby molecular clouds}

\author{Xing-Jian Lv}
\email{lvxj@ihep.ac.cn}
 \affiliation{%
 Key Laboratory of Particle Astrophysics, Institute of High Energy Physics, Chinese Academy of Sciences, Beijing 100049, China}
\affiliation{
 University of Chinese Academy of Sciences, Beijing 100049, China 
}%
 \author{Xiao-Jun Bi}
 \email{bixj@ihep.ac.cn}
\affiliation{%
 Key Laboratory of Particle Astrophysics, Institute of High Energy Physics, Chinese Academy of Sciences, Beijing 100049, China}
\affiliation{
 University of Chinese Academy of Sciences, Beijing 100049, China 
}%
\author{Kun Fang}
\email{fangkun@ihep.ac.cn}
\affiliation{%
 Key Laboratory of Particle Astrophysics, Institute of High Energy Physics, Chinese Academy of Sciences, Beijing 100049, China}
 \author{Peng-Fei Yin}
\email{yinpf@ihep.ac.cn}
\affiliation{%
 Key Laboratory of Particle Astrophysics, Institute of High Energy Physics, Chinese Academy of Sciences, Beijing 100049, China}
\author{Meng-Jie Zhao}
\email{zhaomj@ihep.ac.cn}
 \affiliation{%
 Key Laboratory of Particle Astrophysics, Institute of High Energy Physics, Chinese Academy of Sciences, Beijing 100049, China}
\affiliation{
China Center of Advanced Science and Technology, Beijing 100190, China 
}%



\date{\today}

\begin{abstract}
The observed spectral bump in cosmic-ray (CR) proton and helium spectra, along with the phase and amplitude evolution of CR dipole anisotropy, provide plausible yet indirect evidence for the presence of a nearby CR source. This study investigates the potential of giant molecular clouds (GMCs) located near the solar system to act as natural probes of CRs from the nearby source, with their gamma-ray emissions serving as indicators of spatial variations in CR flux within the solar neighborhood resulting from this source. We show that a nearby source, accounting for the CR data, could imprint distinct features on the $\gamma$-ray spectra of different GMCs. We expect that these features are detectable by LHAASO and upcoming high-energy $\gamma$-ray observatories, providing a powerful test for the hypothesized nearby source. Notably, we find that determining the energy dependence of the $\gamma$-ray spectral index offers a promising approach to investigate the nearby source and constrain its distance. Conversely, if the spectral bump is a widespread Galactic phenomenon, the energy dependence would exhibit uniformity across all GMCs, distinguishing the underlying mechanism accounting for the spectral bump from the scenario involving a nearby source.
\end{abstract}
\maketitle


\section{\label{sec:level1}INTRODUCTION}

\begin{figure}[htbp]
\includegraphics[width=0.43\textwidth]{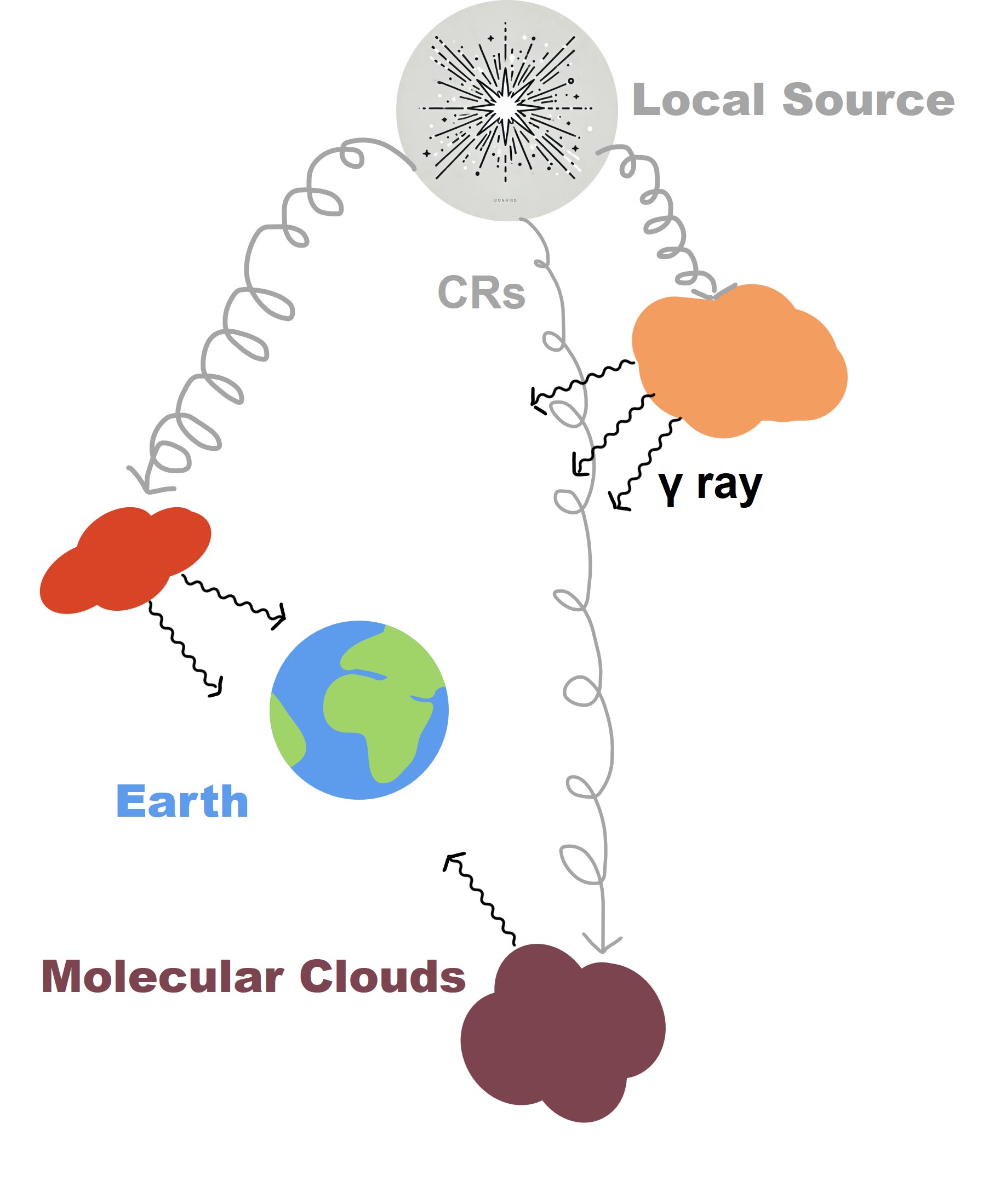}\\
\captionsetup{justification=raggedright}
\caption{Cartoon illustrating the basic concept of our work: GMCs near the solar system can be utilized as detectors for CRs originating from a nearby CR source.  This is achieved through the observation of $\gamma$-ray emissions generated as a result of interactions with CRs at distinct GMCs. GMCs positioned closer to the source receive a stronger flux of CRs compared to those situated at a greater distance, thereby impacting the $\gamma$-ray emission they produce.
\label{fig:schematic}}
\end{figure}



The cosmic-ray (CR) spectrum exhibits deviations from a simple power-law function within the energy range of GeV to PeV, suggesting an expansion of the standard CR scenario~\cite{Gabici:2019jvz}. 
The initial deviation, a spectral hardening at $\sim 200$ GV, has been firmly established through independent observations conducted by various experiments~\cite{Panov:2006kf, Ahn:2010gv, Yoon:2017qjx, PAMELA:2011mvy, AMS:2015azc, AMS:2015tnn}. 
Extending the range of measurements to higher energies, DAMPE has further confirmed the spectral softening at $\sim 10$ TV in both the proton (4.7$\sigma$)~\cite{DAMPE:2019gys} and helium (4.3$\sigma$)~\cite{Alemanno:2021gpb} spectra, thus corroborating earlier indications~\cite{Yoon:2017qjx, Atkin:2018wsp}. This softening, combined with the earlier observed hardening, constitutes the characteristic ``TeV bump". Most recently, at even higher energies, evidences of a new spectral hardening beyond the TeV bump have been observed in the DAMPE $p+\rm He$ spectrum~\cite{DAMPE:2023pjt} and the GRAPES-3 proton spectrum~\cite{GRAPES-32024mhy}.

Meanwhile, CR dipole anisotropy measurements have revealed unexpected features~\cite{Ahlers:2016rox} that intriguingly coincide with the CR spectral bumps. The amplitude of anisotropy shows a rise from $\sim 1$ TeV to $\sim 10$ TeV, followed by a decline up to $\sim 100$ TeV. Furthermore, the phase deviates from the direction of the Galactic center below $\sim 100$ TeV, contrary to the expectation of the conventional diffusion model~\cite{Strong:2007nh}. Instead, it transitions abruptly from R.A. $\sim 4$ hrs towards the Galactic center at $\sim 100$ TeV.

The co-evolution of deviations in the CR energy spectrum and dipole anisotropy from the standard CR scenario can be naturally explained by the contribution from a nearby CR source~\cite{Liu:2018fjy, Yue:2019sxt, Fang:2020cru}.
A nearby source with a cutoff near $\mathcal{O}(10)$ TeV can account for the TeV bump in the CR energy spectra. Below $\sim 10$ TeV, this nearby source predominantly influences the anisotropy, aligning it with the direction of the source. Above $\sim 100$ TeV, the Galactic CR background becomes dominant,  resulting in the anisotropy pointing towards the Galactic center. The features in the anisotropy amplitude arise automatically from the competition between the nearby source and the background CR flux.


Nevertheless, all current indications of a nearby CR source stem from localized CR measurements carried out within the confines of the solar system, providing only indirect support. If such a source indeed exists, the CR flux would exhibit spatial variations in the vicinity of the solar neighborhood, notably intensifying in proximity to the source. The verification of these variations would provide direct evidence for the nearby source.

Giant molecular clouds (GMCs) near the solar system are ideal probes for this investigation. Their $\gamma$-ray emissions, generated by interactions between CRs and GMC gas, can directly trace the CR flux at their respective locations. Previous studies using $\gamma$-rays from GMCs to probe CR properties have often assumed a uniform CR spectral shape for all GMCs near the solar system~\cite{Neronov:2017lqd, Aharonian:2018rob, Albert:2021cwz, Liu:2023phh}, neglecting the potential influence of a nearby CR source.
In reality, GMCs closer to the source experience a more intense CR flux originating from the nearby source, while those situated farther away receive a weaker flux, as illustrated schematically in Fig.~\ref{fig:schematic}. The CR flux at each GMC's position is a superposition of the universal background CR sea and the distinctive contribution from the nearby source. Consequently, significant variations in CR energy spectra exist among GMCs, leading to diverse observed $\gamma$-ray. 

In our study, we begin by utilizing the CR proton and helium spectra, along with the amplitude and phase of the CR dipole anisotropy, to characterize the properties of the hypothesized nearby CR source, which can explain the observed TeV bump. Using these results, we predict the CR fluxes at GMCs near the solar system, considering the diverse distances of these GMCs from the nearby CR source. Subsequently, we calculate the $\gamma$-ray spectra originating from these GMCs. Considering the potential variability in the energy dependence of the $\gamma$-ray spectral indices among different GMCs, we explore the feasibility of utilizing this feature to test the hypothesis of a nearby CR source. The observations from the Large High Altitude Air Shower Observatory (LHAASO) and future high-energy high-energy $\gamma$-ray detectors could provide a promising avenue for examining the presence of a nearby CR source through high-precision $\gamma$-ray observations.

This paper is organized as follows. Section~\ref{sec methology} discusses the utilization of local CR measurements to delineate the properties of the nearby CR source. Section~\ref{sec gmc} introduces the selected GMCs and describes the methodology for calculating their $\gamma$-ray spectra. The predicted $\gamma$-ray fluxes are presented in Section~\ref{sec results}. The findings of this study are summarized in Section~\ref{sec:conclusion}.

\section{Properties of the nearby source}\label{sec methology}
\subsection{Data Sets}
To investigate the properties of the potential nearby CR source, we utilize the CR proton and helium spectra, as well as the amplitude and phase of the CR dipole anisotropy measured by various experiments. The effectiveness of these observables in exploring the nearby CR source is discussed in Section~\ref{sec: fit results}. 

For the proton and helium spectra, we utilize the following data sets:
\begin{itemize}
    \item \textbf{Proton:} AMS-02~\cite{AMS:2021nhj}, DAMPE~\cite{DAMPE:2019gys}, GRAPES-3~\cite{GRAPES-32024mhy}.
    \item \textbf{Helium:} AMS-02~\cite{AMS:2021nhj}, DAMPE~\cite{Alemanno:2021gpb}.
    \item \textbf{$\bm p$+He:} AMS-02~\cite{AMS:2021nhj}, DAMPE~\cite{DAMPE:2023pjt}.
\end{itemize}
In our analysis of the AMS-02 data, we focus on the high-energy data points that coincide with DAMPE measurements to mitigate the impact of factors like solar modulation. A slight inconsistency exists between the AMS-02 and DAMPE helium data, which we attribute to an energy-scale offset, represented as $\tilde{E} = \delta \times E$. Following the methodology of Ref.~\cite{Lv:2024wrs}, we manually rescale the DAMPE helium data using $\delta = 1.037$ and the $p+\rm He$ data using $\delta = 1.029$. The total uncertainties are obtained by combining the systematic and statistical errors in quadrature.

In our analysis of the amplitude and phase of the CR dipole anisotropy, we use the data compilation from Ref.~\cite{Ahlers:2016rox}, supplemented by measurements of LHAASO~\cite{Gao:2023jlz}. These data sets, collected by ground-based experiments, are subject to significant systematic uncertainties. Many experiments report only statistical errors, resulting in noticeable systematic discrepancies between different data sets, often exceeding the reported error bars. 
To ensure consistency among different experiments, we manually rescaled the error bars: increasing the amplitude uncertainties to 35\% and the phase uncertainties to $25^\circ$. For detailed information on the rescaling process, refer to Appendix~\ref{app: rescale}.

\subsection{Calculation Methods and Results\label{sec: calculations}}
\begin{figure*}[htbp]
    \begin{center}
        \subfloat{\includegraphics[width=0.45\textwidth]{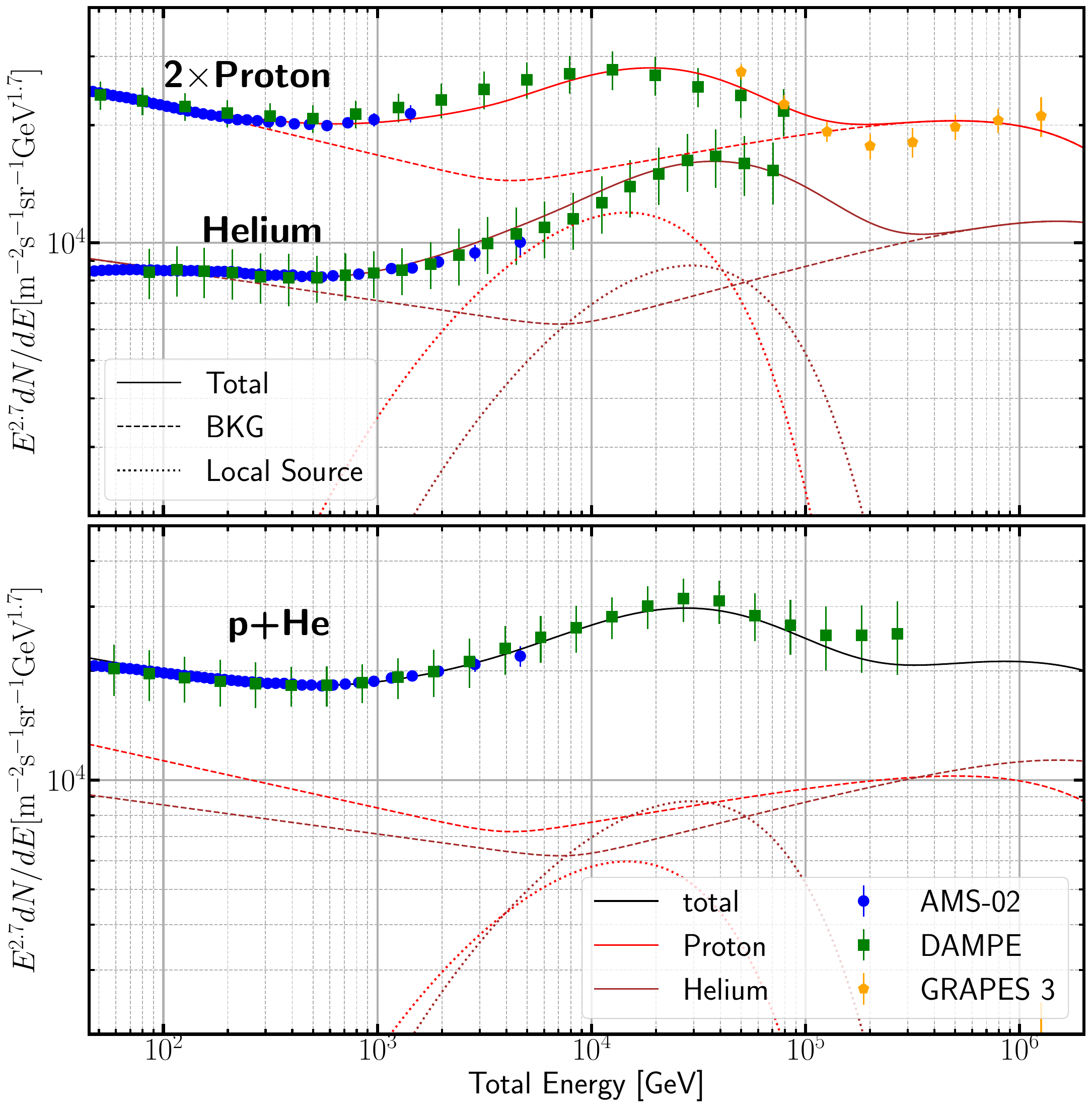}} \hskip 0.03\textwidth
        \subfloat{\includegraphics[width=0.45\textwidth]{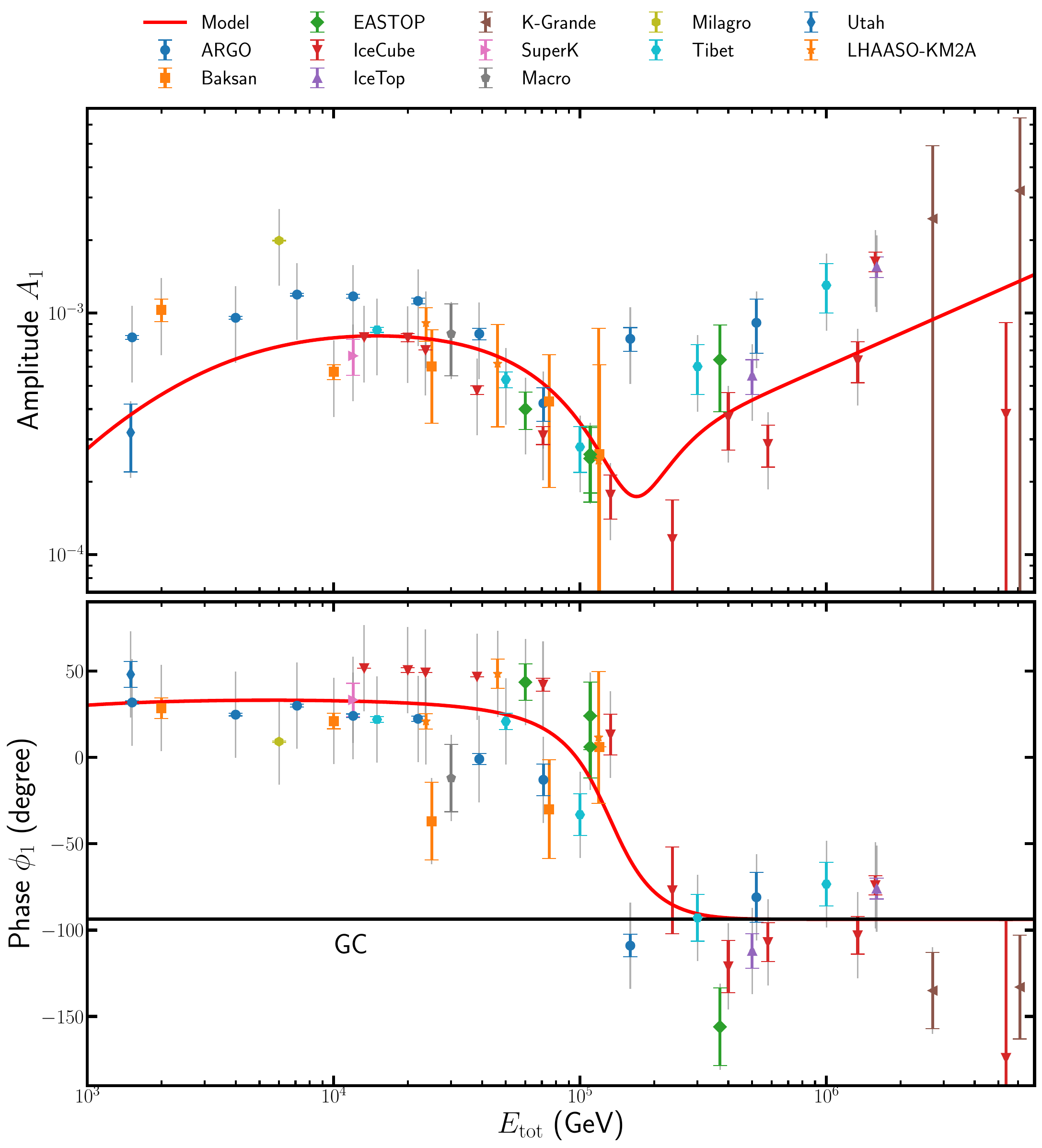}}
    \end{center}
    \captionsetup{justification=raggedright}
    \caption{Left: The proton and helium spectra calculated using the best-fit parameters for the nearby source, compared with data from AMS-02~\cite{AMS:2021nhj}, DAMPE~\cite{DAMPE:2019gys, Alemanno:2021gpb, DAMPE:2023pjt}, and GRAPES-3~\cite{GRAPES-32024mhy}. We fix the source distance in the fitting procedures and test three different cases: $r_s$=100, 250, and 500 pc. The results depicted here correspond to the case of $r_s$ = 250 pc. Right: The best-fit amplitude and phase of the CR dipole anisotropy for $r_s$ = 250 pc, compared with observational data from Ref.~\cite{Ahlers:2016rox} and references therein, as well as LHAASO~\cite{Gao:2023jlz}.}\label{fig: cr fit}
\end{figure*}

\subsubsection{CR spectrum}

The propagation of CR nuclei within our fitting range, $E \gtrsim 50$ GV, is dominated by the diffusion process, with low-energy processes such as solar modulation and reacceleration being negligible. For a nearby CR source, the diffusion equation can be reliably solved within a spherical geometry under the assumption of an infinite boundary condition, as the relevant spatial scales are much smaller than the vertical extent of the diffusion halo (several kpc~\cite{Yuan:2017ozr}). Treating the nearby source as a point source with burst-like CR injection, the differential intensity at Earth, as a function of the total energy $E$, is given by:
\begin{equation}
I_{\mathrm{ls}}\left(E\right)=\frac{c}{(4 \pi)^{5 / 2}\left(D t_s\right)^{3 / 2}} \exp \left(-\frac{r_s^2}{4 D t_s}\right) Q(E)\;,
\label{eqn: ls}
\end{equation}
where $c$ is the speed of light, $r_s$ and $t_s$ denote the distance and age of the nearby source, respectively, and $D(E) = D_0 \left(R / 1 \, \mathrm{GV}\right)^{1/3}$ is the diffusion coefficient, with $R = E / (Z e)$ as the rigidity of a nucleus with charge number $Z$. The injection spectrum, $Q(E)$, is modeled as:
\begin{equation}
Q(E) = Q_0 \left(E / 1 \, \mathrm{GeV}\right)^\gamma \exp(-E / Z E_c)\;,
\end{equation}
where the cutoff scales with $Z$, as expected if the cutoff arises from acceleration or propagation processes~\cite{Hoerandel:2002yg}. The spectral softening at $\sim 10$ TV is naturally explained by this cutoff. The injection spectral index is taken as $\gamma = 2.15$, a typical value for Fermi acceleration~\cite{Erlykin:2001hz}, which is also consistent with $\gamma$-ray and radio observations of supernova remnants~\cite{Caprioli:2011lfx, Bell:2011cs}. An injection energy ratio of 1:1 between protons and helium is assumed, consistent with the properties of the nearby source described in Ref.~\cite{Lv:2024wrs}.

A single power-law model cannot describe the background spectrum, as the spectral index below 200 GV is too soft to account for both the all-particle spectrum at the knee and the new hardening beyond the TeV bump observed in the DAMPE $p+\rm He$ spectrum~\cite{DAMPE:2023pjt} and the GRAPES-3 proton spectrum~\cite{GRAPES-32024mhy}. Therefore, we adopt a smoothly broken power-law for the background spectrum, given by:
\begin{equation}
\begin{aligned}
I_{\mathrm{bkg}}(E) &= I_0 \left(\frac{E}{\mathrm{TeV}}\right)^{-\alpha}\times\\
& \left[1+\left(\frac{E}{ZE_b}\right)^s\right]^{\frac{\Delta \alpha}{s}}\exp(-\frac{E}{Z E_{\mathrm{knee}}})   
\end{aligned}
\end{equation}
where $E_b$ is the break energy, $\Delta \alpha$ is the spectral change, $s$ determines the smoothness of the transition, and $E_{\mathrm{knee}}$ is the cutoff energy. Based on data from $\sim 50$ GV to $\sim 200$ GV, the parameters for the background spectrum before the TeV bump are obtained to be: 
$I^{\mathrm{proton}}_0 = 4.5\times 10^{-2} \:\mathrm{GeV}^{-1} \mathrm{m}^{-2} \mathrm{s}^{-1} \mathrm{sr}^{-1}$, 
$I^{\mathrm{helium}}_0 = 3.4\times 10^{-2} \:\mathrm{GeV}^{-1} \mathrm{m}^{-2} \mathrm{s}^{-1} \mathrm{sr}^{-1}$, 
$\alpha^{\mathrm{proton}} = 2.83$, and $\alpha^{\mathrm{helium}} = 2.78$. For the background spectrum after the TeV bump, parameters derived from DAMPE and GRAPES-3 measurements yield: $E_b = 4$ TeV, $\Delta \alpha = 0.23$, $s = 5.0$, and $E_{\mathrm{knee}} = 5$ PeV. The break in the background spectrum may originate from propagation effects~\cite{Tomassetti:2012ga, Guo:2015csa, Zhao:2021yzf} or the combined contributions of multiple source populations~\cite{abuzayyad2018kneesecondkneecosmicray, Lv:2024wrs}. The origin and parameters of the background spectrum would not significantly affect our discussions on the nearby source. 

\subsubsection{dipole anisotropy}
The CR anisotropy results from the combined contributions of the CR background and the nearby sources. In general, the total dipole anisotropy can be expressed as~\cite{1971ApL9169S}:
\begin{equation}
\Delta = \frac{\sum_i \bar{I}_i \Delta_i \boldsymbol{n}_i \cdot \boldsymbol{n}_{\max}}{\sum_i \bar{I}_i}\;,
\end{equation}
where $i$ represents the $i$-th origin of CRs, including both the background and potential nearby sources, $\bar{I}$ denotes the mean CR intensity for all components, $\boldsymbol{n}$ denotes the direction of the CR origin, and $\boldsymbol{n}_{\max}$ represents the direction of maximum CR intensity. The right ascension of $\boldsymbol{n}_{\max}$ corresponds to the measured phase of the dipole anisotropy. Here we only consider protons and helium components in the analysis, as they dominate the CR flux in the energy range of interest~\cite{Dembinski:2017zsh}.

The anisotropy due to the nearby source, $\Delta_{\mathrm{ls}}$, is determined by its distance and age and can be expressed as $3r_s / (2ct_s)$~\cite{1971ApL9169S}. The anisotropy of the CR background, $\Delta_{\mathrm{bkg}}$, is expected to follow a power-law in energy~\cite{Ahlers:2016rox}, parameterized as $\Delta_{\mathrm{bkg}} = c_1 (E / 1 \, \mathrm{PeV})^{c_2}$, with $c_1 = 0.6$ and $c_2 = 0.45$, consistent with the energy dependence of the diffusion coefficient inferred from the B/C ratio~\cite{Yuan:2017ozr, Genolini:2019ewc}. The location of the nearby source, $\boldsymbol{n}_{\mathrm{ls}}$, is treated as a free parameter under the constraint that it lies on the Galactic plane, while the direction of the background anisotropy, $\boldsymbol{n}_{\mathrm{bkg}}$, points toward the Galactic center.

\subsection{Fitting results}\label{sec: fit results}
\begin{table}[ht]
    \centering
    \renewcommand{\arraystretch}{1.5}
    \captionsetup{justification=raggedright}
    \caption{The fitted values of the parameters in our nearby source model, accounting for the AMS-02, DAMPE, and GRAPES-3 spectra, along with the dipole CR anisotropy data.}\label{tab: results}
    \begin{tabular}{lccc}
        \toprule
        \midrule
        Parameter & {$r_s = 100$ pc} & {$r_s = 250$ pc} & {$r_s = 400$ pc} \\
        \midrule
        $D_0 \, (10^{26} \, \mathrm{cm^2 \, s^{-1}})$ & $1.09\pm0.08$ & $2.73\pm0.20$ & $4.37\pm0.33	$ \\
        $t_s \, (100 \, \mathrm{kyr})$ & $2.86\pm0.19$ & $7.2	\pm0.5$ &$11.5\pm0.8$ \\
        $E_c \, (\mathrm{TeV})$ & $38.7\pm2.8$ & $38.7\pm2.8$ & $38.7\pm2.8$ \\
        $\mathcal{E}_\mathrm{cr} \, (10^{50} \, \mathrm{erg})$ & $0.21\pm0.01$ & $3.31\pm0.014$ & $13.56\pm0.59$\\
        RA (degree) & $37\pm4$ & $37\pm4$& $37\pm4$\\
        $\chi^2$/Dof & 259/270 & 259/270 & 259/270\\
        \bottomrule
    \end{tabular}
\end{table}
In our nearby source model, the parameters to be determined include $D_0$, $r_s$, $t_s$, $E_c$, $\mathrm{RA}$, and $\mathcal{E}_\mathrm{cr}$. Assuming that the nearby source is situated in the Galactic plane, the right ascension of the source, denoted as RA, is considered as a free parameter. The parameter $\mathcal{E}_\mathrm{cr}$ represents the energy injected into CRs, which determines the normalization $Q_0$ of the injection spectrum.

In principle, the local CR data under consideration can effectively break the degeneracies among the parameters within our nearby source model, once the distance to the source, $r_s$, is specified. The bump structure in the dipole amplitude is sensitive to the ratio $r_s / t_s$, enabling the determination of $t_s$ for a given $r_s$. Furthermore, the suppression of the nearby source's contribution to the CR spectrum at low energies depends on the combination $r_s^2 / (D_0 t_s)$, as shown in Eq.(\ref{eqn: ls}), facilitating the inference of $D_0$. Note that we do not adopt the Galactic average diffusion coefficient inferred from the B/C ratio, as the diffusion coefficient in the solar neighborhood may deviate from this value \cite{Fang:2020cru}. Moreover, the height of the TeV bump in the CR spectrum can determine $\mathcal{E}_\mathrm{cr}$, while the CR dipole phase provides constraints on $\mathrm{RA}$. Consequently, the properties of the nearby source in our model, as inferred from the local CR data, are primarily dependent on the source distance rs $r_s$. In our analysis, we maintain $r_s$ fixed and and explore three cases: $r_s = 100$, 250, and $400 \, \mathrm{pc}$.

The results of the fitting for five parameters are summarized in Table~\ref{tab: results}, while the best-fit CR spectrum and CR anisotropy for the case of $r_s = 250 \, \mathrm{pc}$ are shown in Fig.~\ref{fig: cr fit} as an example. 
The required energy injected into CRs is positively correlated with the source distance. Assuming a typical supernova explosion energy of $10^{51} \mathrm{erg}$ and a CR conversion efficiency of $\sim 10\%$, the expected energy injected into CR is $\mathcal{E}_\mathrm{cr} \sim 10^{50} \mathrm{erg}$~\cite{Blasi:2013rva}. As shown in Table~\ref{tab: results}, the required CR energies are approximately 0.2, 3.0, and 10 times the typical value for source distances of $100 \, \mathrm{pc}$, $250 \, \mathrm{pc}$, and $400 \, \mathrm{pc}$, respectively. This suggests that the nearby source distance should lie within the range of $100$–$400 \, \mathrm{pc}$, as distances outside this range would lead to significant deviations from the typical CR injection energy value.

\section{Cosmic-ray interaction in molecular clouds}\label{sec gmc}
\begin{figure}[t]
\includegraphics[width=0.45\textwidth]{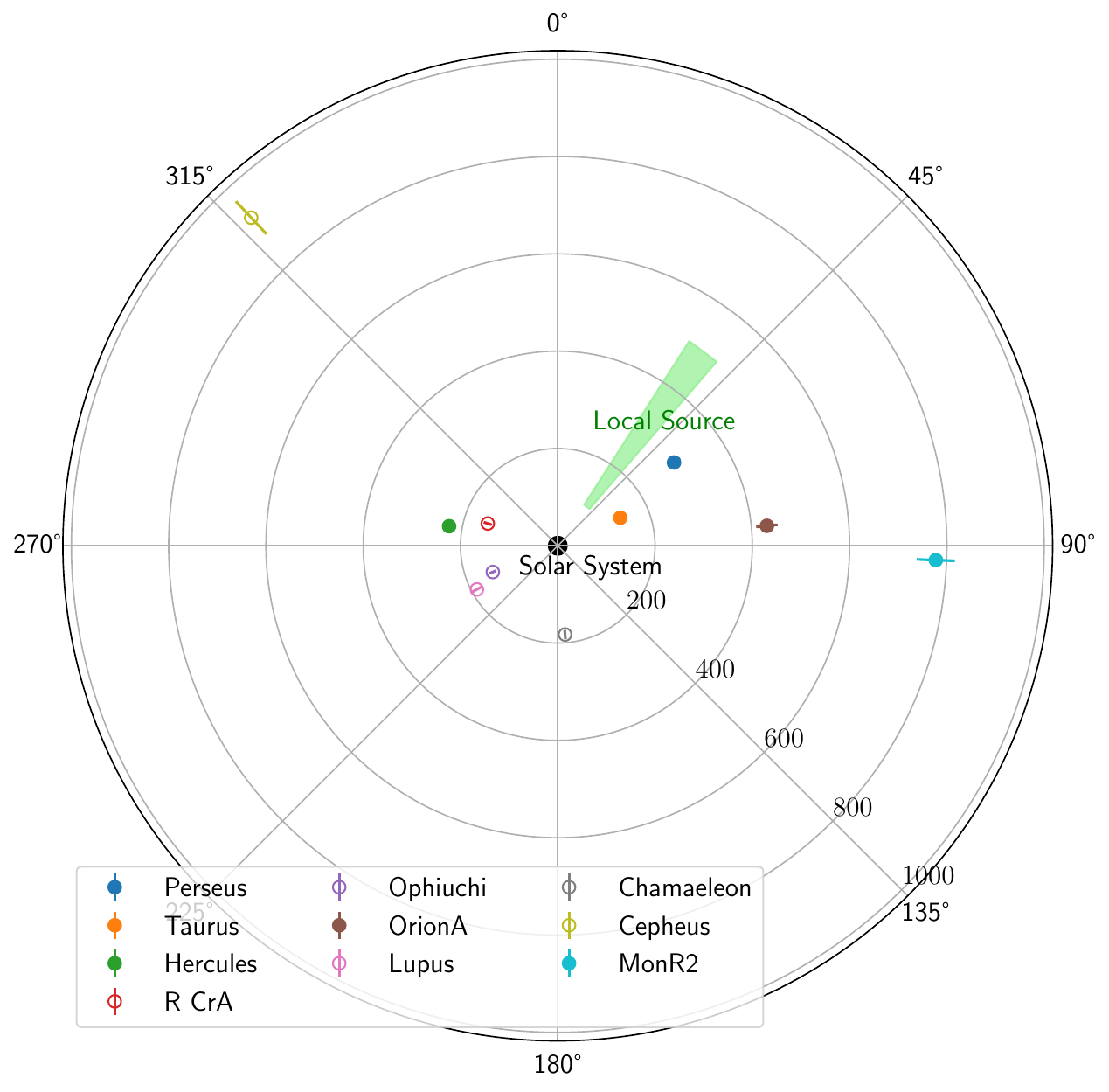}\\
\captionsetup{justification=raggedright}
\caption{The positions of the selected GMCs in the celestial equatorial plane. The positions and their relative uncertainties are taken from Ref.~\cite{zuckerLargeCatalogAccurate2019}. Solid points represent GMCs within the LHAASO's field of view, while hollow points represent other selected GMCs. The green band illustrates the fitted direction of the nearby source for $r_s = 100$, 250, and 400 pc.}\label{fig: gmc}
\end{figure}
The inelastic interactions of CRs with gas in GMCs result in the production and subsequent decay of secondary unstable particles, primarily $\pi$-mesons. The decay of these particles generates $\gamma$-rays, neutrinos, and electrons. Assuming that CRs can freely penetrate the cloud, the $\gamma$-ray flux from a passive GMC depends on a characteristic parameter of the GMC, denoted as $A \equiv M / d^2$~\cite{Aharonian:2000zd}, where $M$ is the GMC's mass and $d$ is the distance to the GMC. The $\gamma$-ray spectrum can be given by
\begin{equation}
F_\gamma\left(E_\gamma\right)=\xi_{\mathrm{N}} \frac{A}{m_p} \int \frac{d \sigma_{\mathrm{pp} \rightarrow \gamma}\left(E_{\mathrm{p}}, E_\gamma\right)}{d E_\gamma} I\left(E_{\mathrm{p}}\right) d E_{\mathrm{p}}\;,
\end{equation}
where $d \sigma_{\mathrm{pp} \rightarrow \gamma}\left(E_{\mathrm{p}}, E_\gamma\right)/d E_\gamma$ represents the differential cross-section for proton-proton interactions leading to gamma-ray production, as calculated in Ref.~\cite{Kafexhiu:2014cua}, $m_p$ represents the proton mass, $I\left(E_{\mathrm{p}}\right)$ denotes the proton flux at the GMC location, and 
$\xi_{\mathrm{N}}$ denotes the nuclear enhancement factor accounting for the contribution of nuclei to the $\gamma$-ray flux. Based on previous studies~\cite{Aharonian:2018rob, Albert:2021cwz}, $\xi_{\mathrm{N}}$ is taken to be 1.8. The proton flux $I\left(E_{\mathrm{p}}\right)$ includes contributions from both the background CR sea and the nearby source. While the background CR flux remains consistent across all GMCs, the contribution from the nearby source increases as the GMC approaches the source.
\begin{table}[ht]
    \centering
    \renewcommand{\arraystretch}{1.5}
    \captionsetup{justification=raggedright}
    \caption{The properties of the selected GMCs: calculated $A$ factor, equatorial coordinates (Dec, RA), and distances to the nearby source for $r_s = 100$, 250, and 400 pc. The values of the $A$ factor, taken from Refs.\cite{Aharonian:2018rob, Albert:2021cwz}, are normalized as $A = M_5/d^2_{\mathrm{kpc}}$, where $M_5 = M / 10^5 M_{\odot}$ and $d_{\mathrm{kpc}} = d / 1 \mathrm{kpc}$. The GMC positions are based on data from Ref.~\cite{zuckerLargeCatalogAccurate2019}. GMCs marked with an asterisk are located within the LHAASO field of view.}\label{tab: gmc}
    \begin{tabular}{lcccccc}
        \toprule
        \midrule
        GMC & {$A$} & {Dec.[$^\circ$]} & {RA.[$^\circ$]} & {$d_{100}$[pc]} & {$d_{250}$[pc]} & {$d_{400}$[pc]} \\
        \midrule
        $\mathrm{Perseus}^*$ & 1.55 & 31.83 & 54.39 & 215&152& 212 \\
        $\mathrm{Taurus}^*$ & 5.64 & 26.49 & 65.89 &90&167& 305 \\
        $\mathrm{Hercules}^*$ & 1.16 & 14.23 & 280.20 &248&337& 459 \\
        R CrA & 0.63 & -37.02 & 287.64 &229&367& 511 \\
        Ophiuchi & 3.94 & -24.33 & 248.01 &228&369& 516 \\
        $\mathrm{OrionA}^*$ & 6.76 & -7.12 & 84.56 &422&450& 521 \\
        Lupus & 1.00 & -35.86 & 241.66 &280&425& 572 \\
        Chamaeleon & 2.98 & -77.33 & 175.28 &278&426& 574 \\
        Cepheus & 2.52 & 70.47 & 316.93 &839&722& 619 \\
        $\mathrm{MonR2}^*$ & 0.87 & -6.60 & 92.17 &767&774& 810 \\
        \midrule
        \bottomrule
    \end{tabular}
\end{table}

For the target GMCs, we select those located within 1 kpc of the solar system that have been used in previous studies~\cite{Neronov:2017lqd, Aharonian:2018rob, Albert:2021cwz}. However, these studies assume that the CR spectral shape is identical for all GMCs near the solar system, thereby neglecting the possibility of a nearby source contributing to the CR fluxes. The values of the $A$ factor for these GMCs are also adopted from these works. A total of 10 GMCs have been selected. Their properties, along with their distances to the nearby source for $r_s = 100$, 250, and 400 pc, are listed in Table~\ref{tab: gmc}. The locations of these GMCs  relative to the solar system based on data from Ref.~\cite{zuckerLargeCatalogAccurate2019} and the fitted direction of the nearby source are illustrated in Fig.~\ref{fig: gmc}.

\section{Expected gamma-ray flux}\label{sec results}
\begin{figure*}[htbp]
    \begin{center}
        \subfloat{\includegraphics[width=0.80\textwidth]{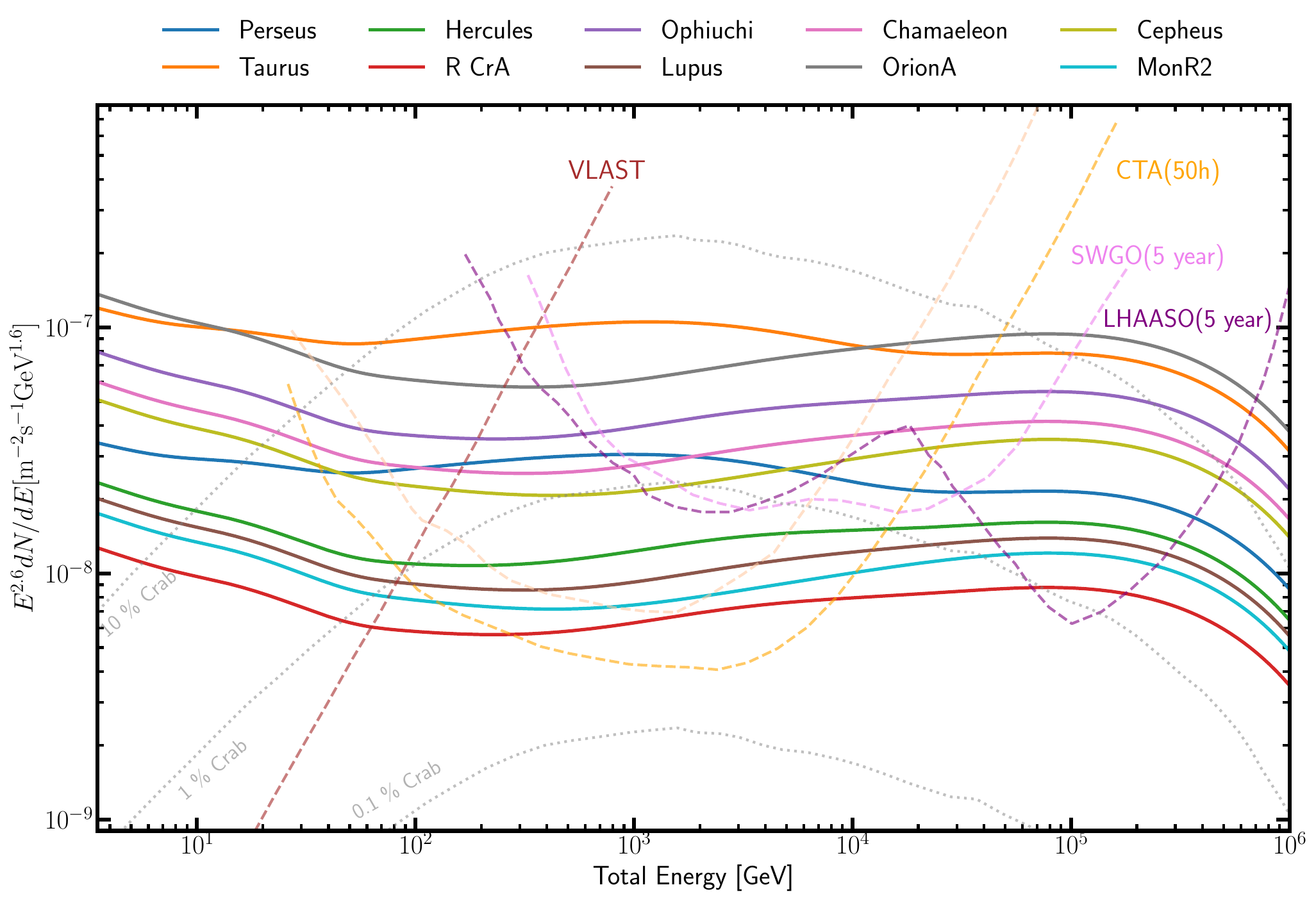}}\\
        \subfloat{\includegraphics[width=0.45\textwidth]{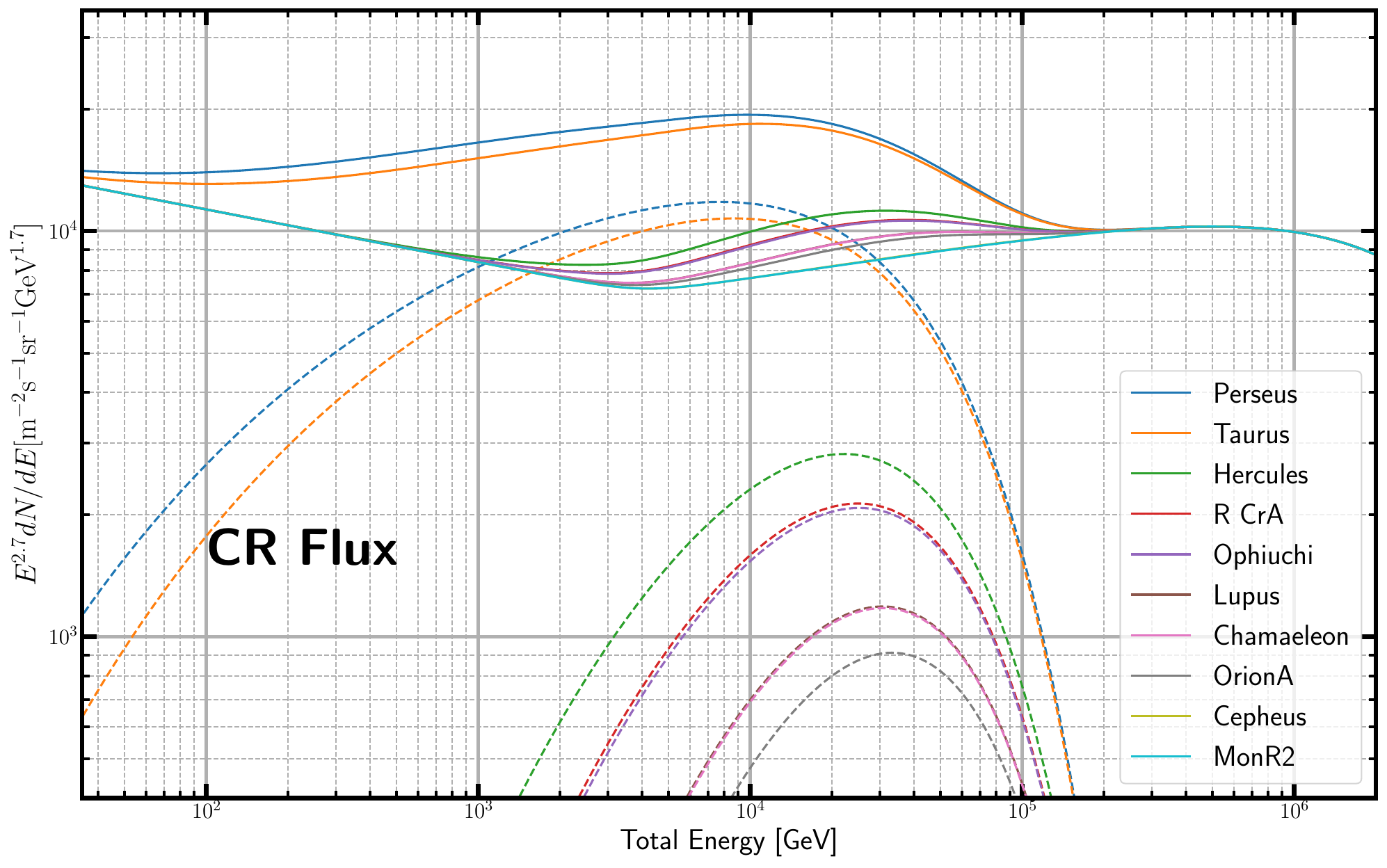}} \hskip 0.03\textwidth
        \subfloat{\includegraphics[width=0.45\textwidth]{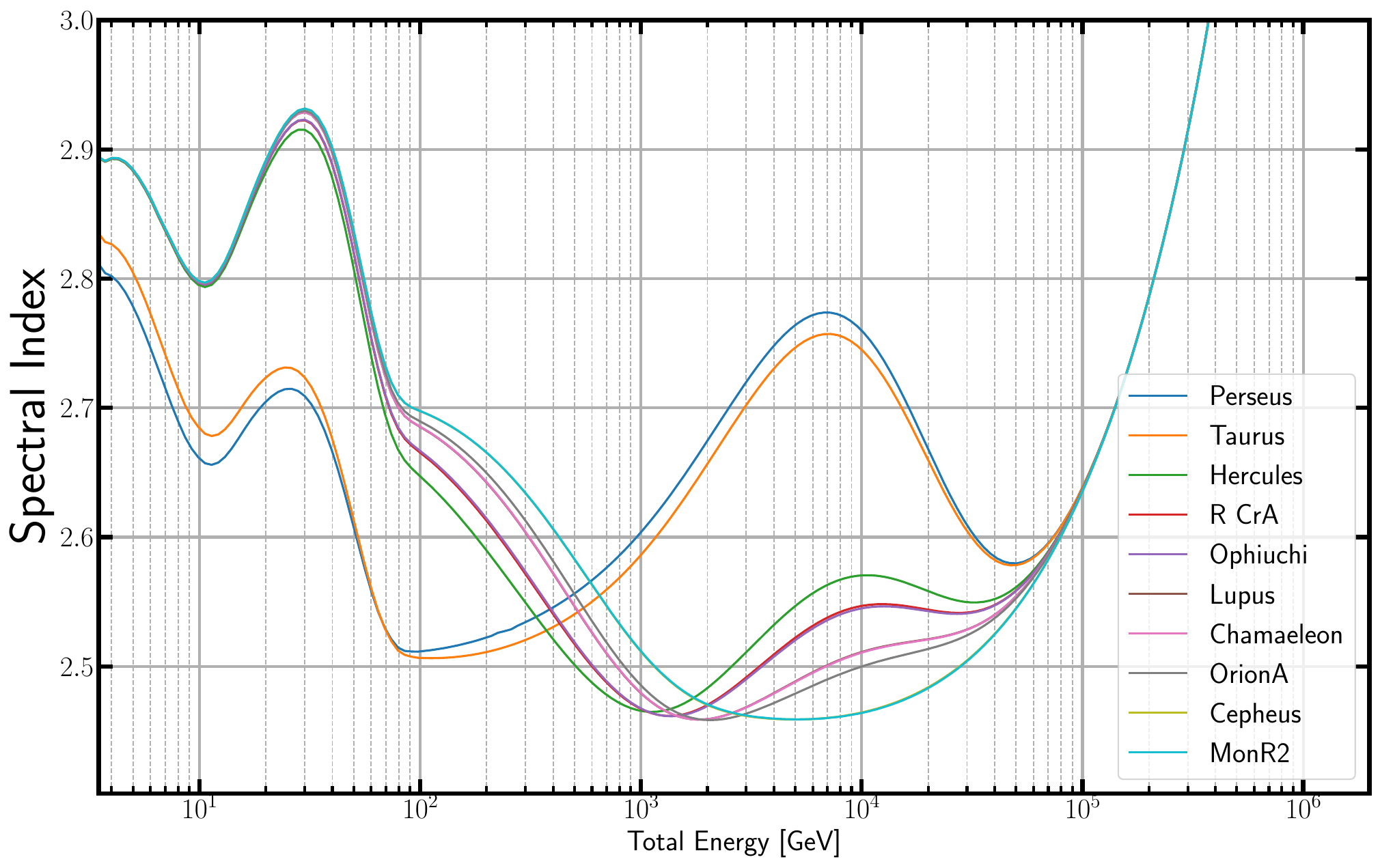}}
    \end{center}
    \captionsetup{justification=raggedright}
    \caption{Top: The expected $\gamma$-ray spectrum originating from selected GMCs a source distance of $r_s = 250$ pc, compared to the point-like source sensitivities of currently operating and future $\gamma$-ray instruments, taken from Ref.~\cite{Peron:2021uft}. The brown curve represents the sensitivity of a hypothetical Very Large Area Telescope (VLAT) with an effective area three times larger than Fermi-LAT. The orange and light orange curves correspond to the expected sensitivity of CTA from the southern and northern sites, respectively, for 50 hours of observations. The violet curve indicates the sensitivity of SWGO after five years of observations. The purple curve represents the sensitivity of LHAASO for five years of observations. Bottom: On the left side, the expected proton spectra at the locations of the GMCs are shown. The solid lines represent the total proton flux, while the dashed lines specifically show the contribution from the nearby source. On the right side, the expected energy dependency of the spectral index of the $\gamma$-ray spectrum for various GMCs is displayed.}\label{fig: gamma flux}
\end{figure*}
The main results of this work are presented in the top panel of Fig.~\ref{fig: gamma flux}, showing the expected $\gamma$-ray flux from several major GMCs near the solar system for a source distance of $r_s = 250$ pc. These flux predictions are compared with the sensitivities of current and future $\gamma$-ray detectors, calculated for their typical exposure times.
The sensitivity curves, taken from Ref.~\cite{Peron:2021uft}, include both ground-based and space-borne detectors. Ground-based instruments include the  currently operational LHAASO~\cite{LHAASO:2019qtb}, the proposed Southern Wide-field Gamma-ray Observatory (SWGO~\cite{Albert:2019afb}), and the Cherenkov Telescope Array (CTA~\cite{CTAConsortium:2017dvg}). Additionally, a hypothetical space-borne detector, denoted as $\mathrm{VLAT}-\Sigma = 3$, represents the sensitivity of a Very Large Area Telescope (VLAT) with an effective area three times larger than that of Fermi-LAT~\cite{Fermi-LAT:2011iqa}. Examples of proposed instruments with similar capabilities include the Very Large Area gamma-ray Space Telescope (VLAST~\cite{Pan:2024adp}) and the High Energy cosmic-Radiation Detection (HERE~\cite{Cagnoli:2024hgd}).
For completeness, the expected $\gamma$-ray fluxes for source distances of $r_s = 100$ and $400$ pc are provided in Appendix~\ref{app: rs}.

The distance from the nearby source significantly impacts the CR flux at the GMC's location, thereby influencing the $\gamma$-ray flux they emit. This influence manifests in two key ways:
\begin{enumerate}
  \item Increased flux with proximity: The closer a GMC is to the nearby source, the higher the CR flux it receives from the source. This results in a more pronounced bump structure in its local CR energy spectrum due to the enhanced contribution from the nearby source.
  \item Earlier spectral hardening for closer GMCs: The energy dependence of the diffusion coefficient causes low-energy CRs to diffuse at a slower rate, requiring more time for them to transport longer distances. Consequently, GMCs in closer proximity to the nearby source are influenced by its CR flux at lower energies, leading to an earlier hardening in their CR spectrum.
\end{enumerate}
These effects are clearly illustrated in the lower-left panel of Fig.~\ref{fig: gamma flux},  illustrating the anticipated CR flux at various GMCs. The two closed GMCs to the nearby source, namely Perseus and Taurus, exhibit a pronounced bump in their CR spectrum along with an earlier onset of hardening. In contrast, other GMCs located at much greater distances from the nearby source have CR spectra that closely resemble the background spectrum, with spectral hardening occurring at much higher energies. The spectral hardening in these distant GMCs is primarily attributed to the intrinsic hardening of the CR background itself, with only minor contributions from the nearby source.

The variations in the CR energy spectra among different GMCs are naturally reflected in their corresponding $\gamma$-ray spectra, as illustrated in the top panel of Fig.~\ref{fig: gamma flux}. Since the absolute normalization of the $\gamma$-ray flux depends on the $A$ factor of each GMC, which has inherent uncertainties of about 30\%~\cite{Aharonian:2018rob}, it may not be particularly effective in distinguishing between GMCs that are closer or farther from the nearby source. However, the energy dependence of the $\gamma$-ray spectrum remains unaffected by the specific properties of GMCs. Instead, it can serve as a direct probe of the nearby source's contribution to the CR flux at each GMC's location.

The two GMCs closest to the nearby source, Perseus and Taurus, also show a more pronounced bump in their $\gamma$-ray spectra and undergo earlier hardening compared to more distant GMCs. This distinct is evident in the lower-right panel of Fig.~\ref{fig: gamma flux}, which presents the energy dependence of the spectral index of the $\gamma$-ray flux for various GMCs. For Perseus and Taurus, the spectral indices reach their minimum values (corresponding to their spectra at the hardest) before 100 GeV. In contrast, other GMCs, located farther from the nearby source, achieve this minimum at much higher energies, exceeding 1 TeV, more than an order of magnitude later.
CTA is particularly well-suited to cover this energy range, owing to its full sky coverage and excellent energy resolution ($\sim 7\%$~\cite{cherenkov_telescope_array_observatory_2021_5499840}). This capability enables precise measurements of the spectral indices, making it an ideal instrument to distinguish the influence of a nearby CR source on the $\gamma$-ray spectra of GMCs.

LHAASO, with an exceptional sensitivity, is expected to detect many GMCs within its field of view, as denoted by solid points in Fig.~\ref{fig: gmc}. Among these, Taurus (close to the nearby source) and Orion A (far from the nearby source) stand out due to their large $A$ factors. These GMCs can be observed by LHAASO across a broad energy range, enabling the  investigation of spectral index variations with energy. If the observed results match our predictions, the consistency would serve as direct evidence of a nearby CR source. However, caution is warranted, as nearby GMCs, with their significant extensions ($\sim 1^\circ$), experience sensitivity reductions. The reduction factor is determined by a factor of $\sqrt{1 + (\theta/\sigma_{\mathrm{PSF}}(E))^2}$, where $\sigma_{\mathrm{PSF}}$ represents the point spread function of the instrument.

Furthermore, if the TeV bump is a widespread phenomenon  throughout the Milky Way, all selected GMCs would exhibit identical energy dependencies in the spectral index of their $\gamma$-ray flux. Such uniformity would contradict the hypothesis of a nearby CR source, highlighting the importance of investigating the underlying mechanisms shaping the observed CR and $\gamma$-ray spectra.

\section{SUMMARY\label{sec:conclusion}}

The presence of a nearby CR source offers a plausible explanation for the TeV bump evident in the CR proton and helium energy spectra, as well as for the energy evolution of both the amplitude and phase of the dipole anisotropies. However, these investigations are primarily based on local CR measurements within the solar system, offering only indirect evidence for the existence of such a nearby source. 

In this study, we utilize nearby GMCs as probes of the nearby CR source, as these clouds emit $\gamma$-rays that can directly reveal the CR energy spectra at various positions.
We introduce a nearby source to reproduce the measured CR proton and helium spectra, as well as the amplitudes and phases of the CR dipole anisotropy. Subsequently, we predict the $\gamma$-ray flux from a set of selected GMCs near the solar system. 

Our results suggest that many of these chosen GMCs are likely to be detectable by LHAASO and future $\gamma$-ray observatories. We find that the energy dependence of the spectral index of the $\gamma$-ray flux for different GMCs can provide a valuable tool for testing the hypothesized nearby source, with this dependence being significantly influenced by the distances between the GMCs and the source. If the expected differences in the $\gamma$-ray spectra across various GMCs are confirmed through observations, it would serve as compelling evidence supporting the existence of a nearby CR source.


\acknowledgments
This work is supported by 
the National Natural Science Foundation of China under Grants No. 12105292, No. 12175248, and No. 12393853. 

\bibliography{apssamp}
\appendix
\section{Rescaling of the error bars of CR data}\label{app: rescale}
\begin{figure}[t]
\includegraphics[width=0.48\textwidth]{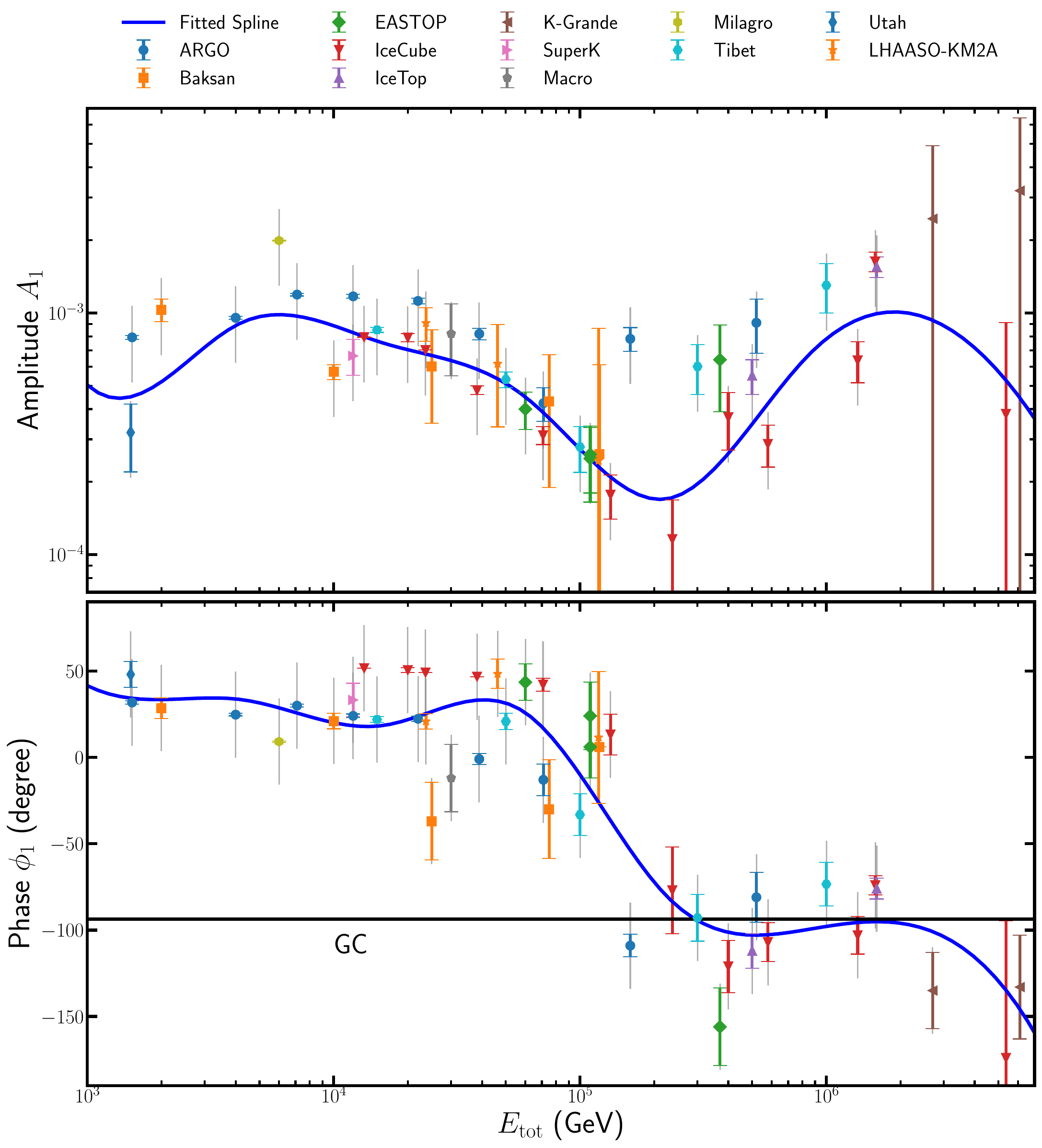}\\
\captionsetup{justification=raggedright}
\caption{The B-spline fitting results for the amplitude (top) and phase (bottom) of the dipole anisotropy, compared with the observational data.
\label{fig: spline}}
\end{figure}
To ensure the consistent combination of data from different experiments without excessively reducing the constraining power, we adopt a strategy to determine the appropriate rescaling of error bars. This strategy involves incrementally increasing the error bars and fitting a model-independent smooth curve through the data points at each step. The rescaling factor is adopted once the $\chi^2$ per data point falls below one.

In constructing the smooth curve, we utilize a cubic B-spline, which is formulated as a linear combination of standard B-splines:
\begin{equation}
f(E)=\sum_k a_{k} \times b_k(\ln (E / \mathrm{GeV}))\;,
\end{equation}
where $f(E)$ represents the amplitude or phase of the dipole anisotropy, $b_k(\ln (E / \mathrm{GeV}))$ are standard cubic B-splines defined over a common vector of knots, and $a_k$ are the coefficients determined through a global fit to all experimental data. The knot locations are arranged on a regular grid in $\ln (E / \mathrm{GeV})$, with a step size around 0.5.

Following this procedure, we obtain rescaled error bars for the dipole amplitude and phase that lead to reasonable $\chi^2$ values. 
Specifically, an error bar of 35\%
for the dipole amplitude leads to 
$\chi^2 = 53$ for 47 data points, and an error bar of $25^\circ$ for the dipole phase results in $\chi^2 = 50$ for 47 data points. We show the fitted B-spline curve compared with the data in Fig.~\ref{fig: spline}. As can be seen, the fitted B-spline provides a model-independent smooth representation of the data.

\section{Different distances of the nearby source}\label{app: rs}
\begin{figure*}[t]
    \begin{center}
        \subfloat{\includegraphics[width=0.45\textwidth]{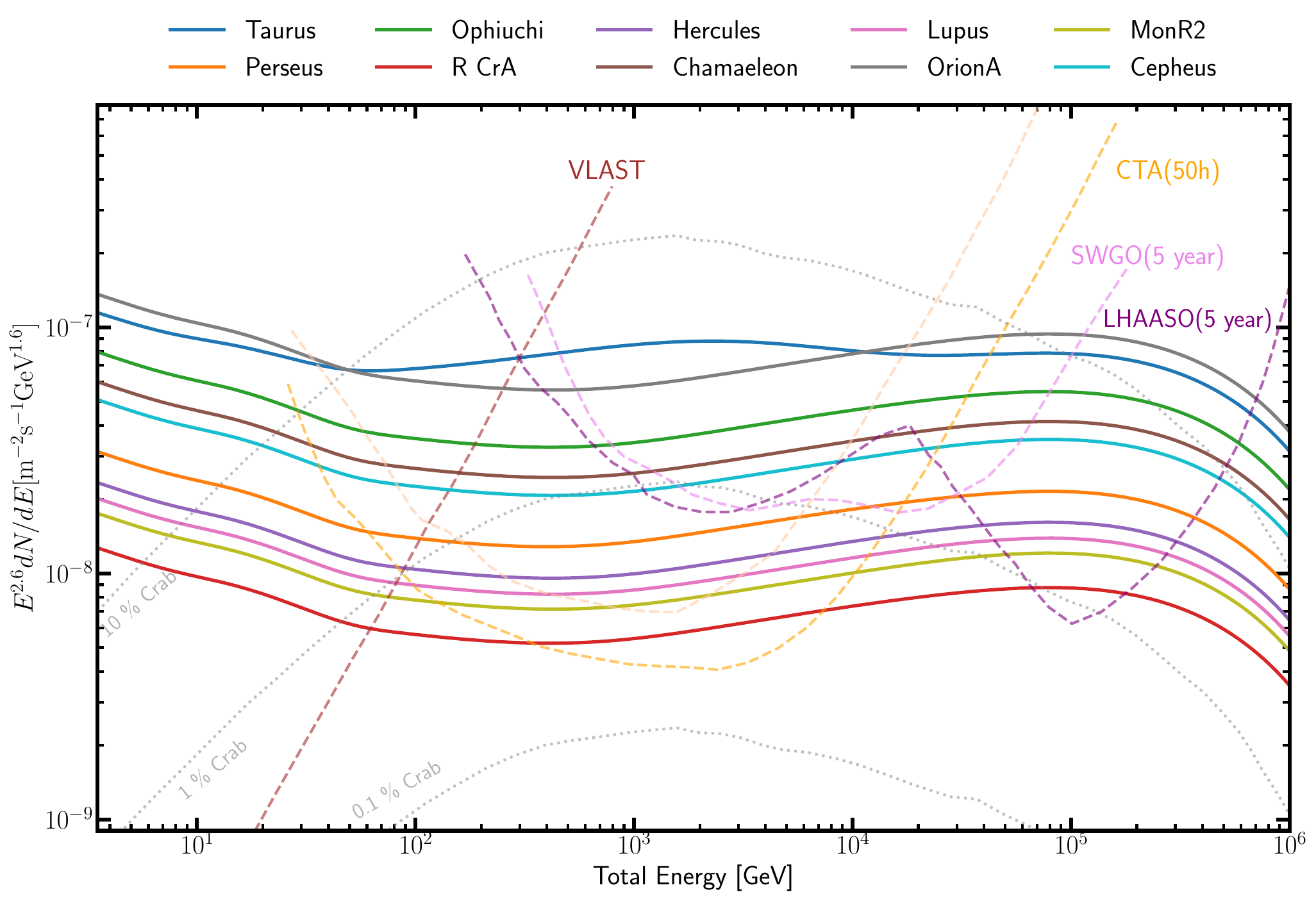}} \hskip 0.03\textwidth
        \subfloat{\includegraphics[width=0.45\textwidth]{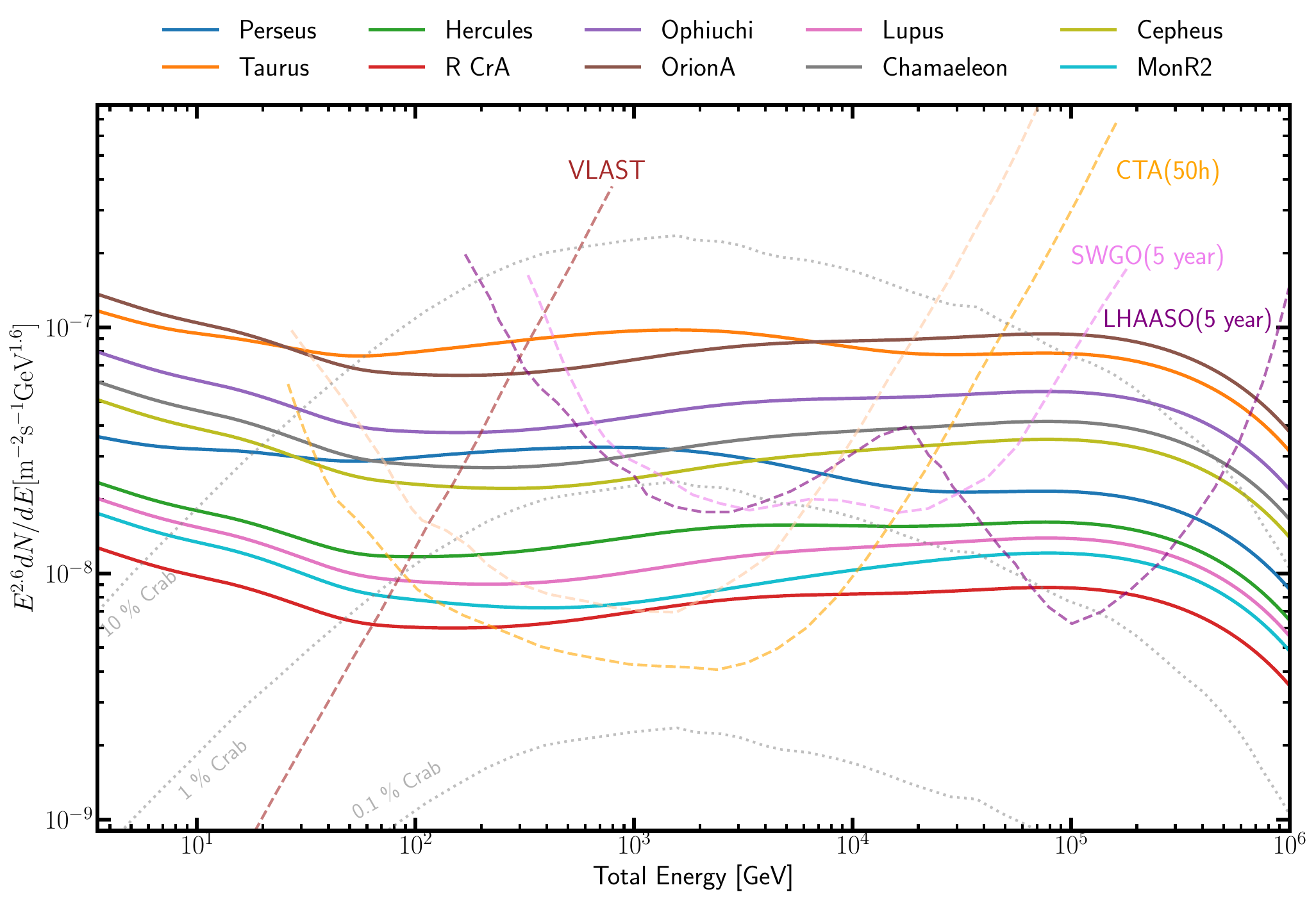}}\\
        \subfloat{\includegraphics[width=0.45\textwidth]{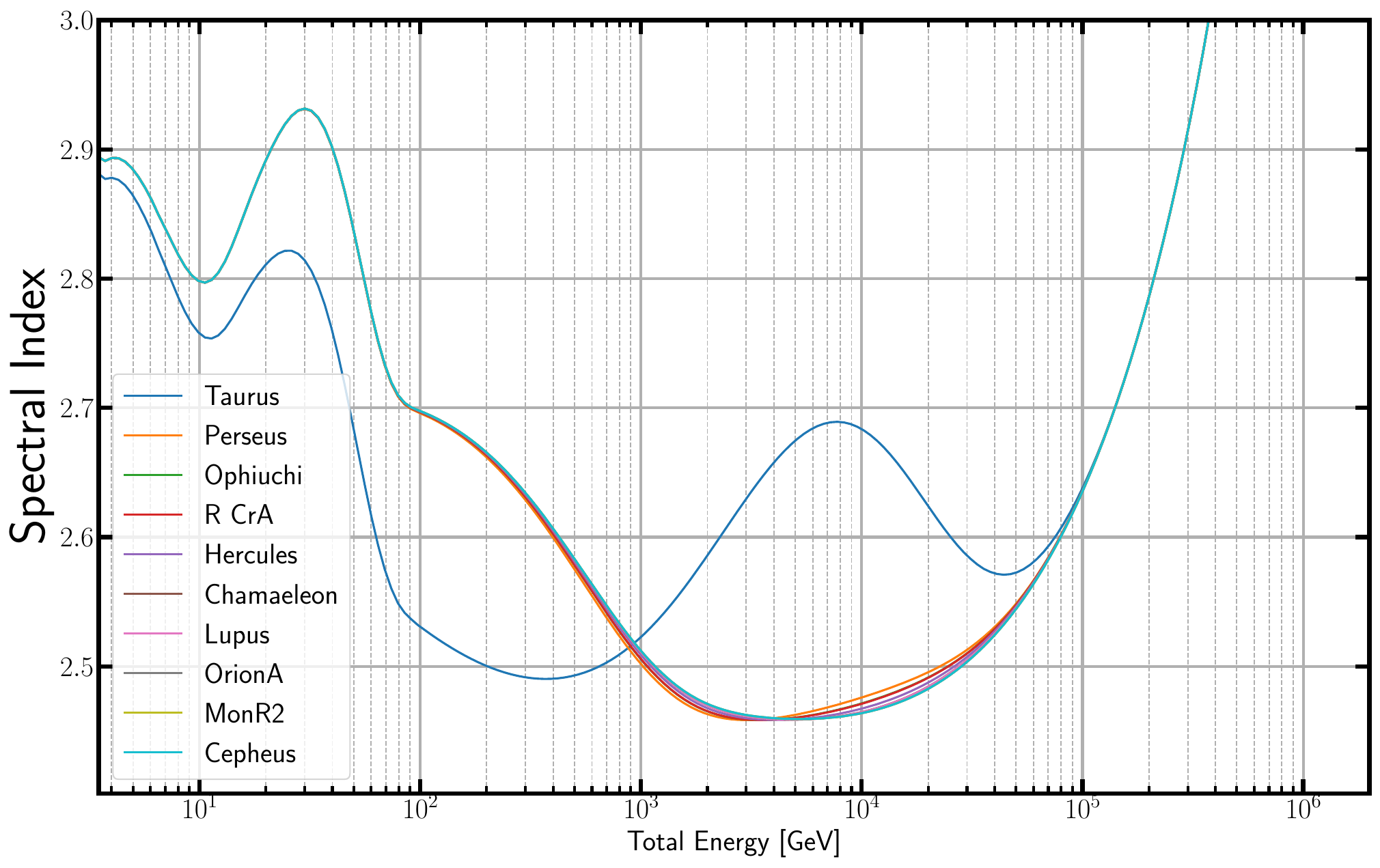}} \hskip 0.03\textwidth
        \subfloat{\includegraphics[width=0.45\textwidth]{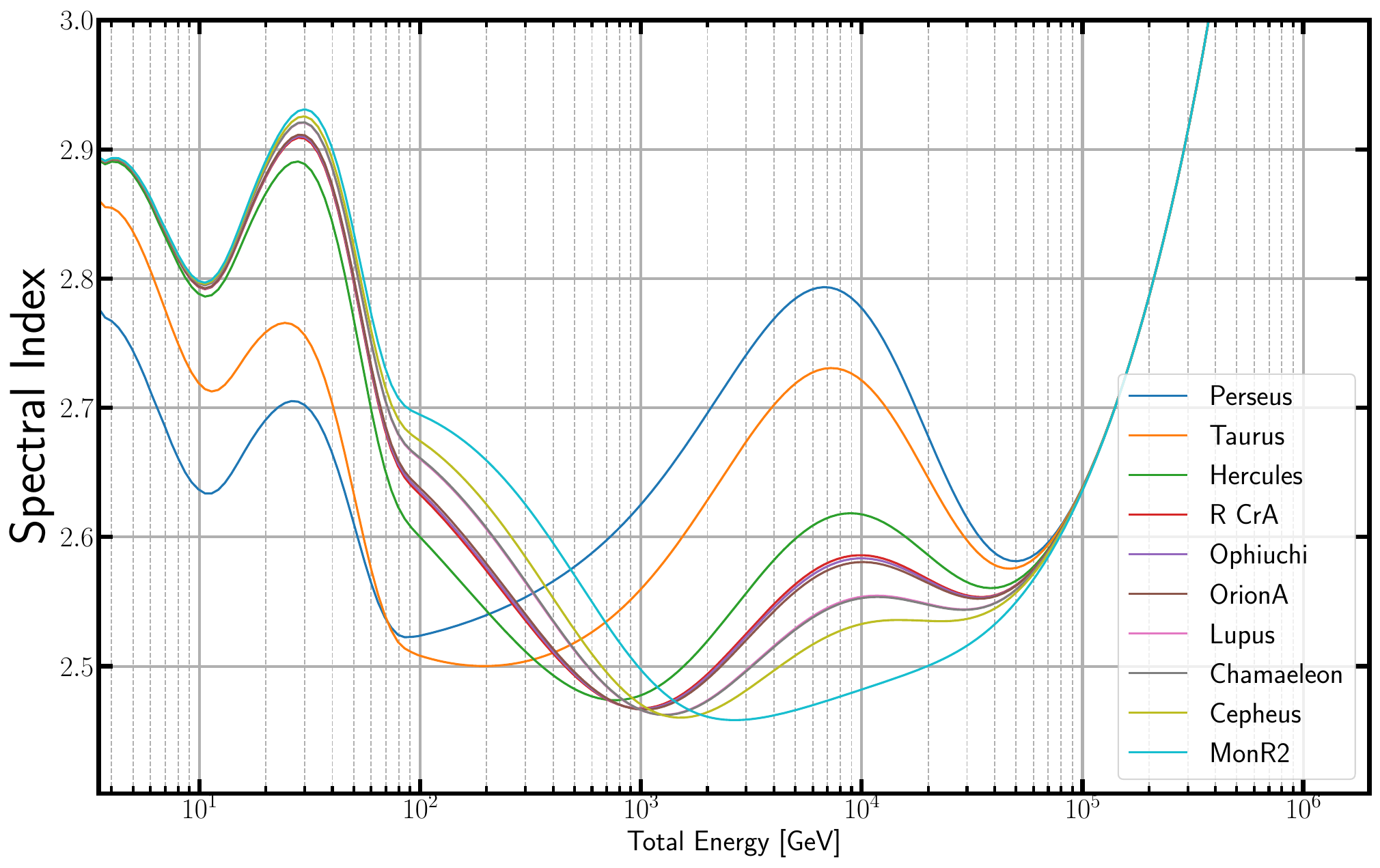}}
    \end{center}
    \captionsetup{justification=raggedright}
    \caption{Top: The expected $\gamma$-ray spectra from selected GMCs for $r_s = 100$ pc (left) and $r_s = 400$ pc (right), compared to the point-like source sensitivities of current and future $\gamma$-ray instruments. Bottom: The expected energy dependence of the spectral index of the $\gamma$-ray flux for different GMCs for $r_s = 100$ pc (left) and $r_s = 400$ pc (right). The meanings of the curves in this figure are the same as that in Fig.~\ref{fig: gamma flux}. The styles of the curves in this figure align with those described in Fig.~\ref{fig: gamma flux}.
}\label{fig: diff rs}
\end{figure*}
In Fig.~\ref{fig: diff rs}, we illustrate the expected $\gamma$-ray spectrum from selected GMCs for two nearby source distances, $r_s = 100$ and $400$ pc. Furthermore, we also show the corresponding energy dependence of the spectral index of the $\gamma$-ray spectrum.
For the case $r_s = 400$ pc, the expected $\gamma$-ray flux is similar to the case of $r_s = 250$ pc discussed in the main text. Notably, Perseus and Taurus, which are in close proximity to the nearby source, display distinct spectral indices compared to the other GMCs. 

In contrast, in the case of $r_s = 100$ pc, only Taurus is sufficiently close to the nearby source to be significantly influenced by it. Despite Perseus being located at a distance of 215 pc from the nearby source in this case (closer than its distance in the $r_s = 250$ and $400$ pc scenarios), the total power injected into CRs by the nearby source is much smaller compared to the other two cases. As a result, even at 215 pc, Perseus is situated too far from the nearby source to experience a significant impact in this case. With future high-precision measurements, the extent to which each GMC is influenced by the nearby source could potentially be used to break the degeneracy in determining the nearby source's distance, $r_s$.



%
\end{document}